\documentclass[amsmath, prd, nofootinbib, showpacs]{revtex4}

\usepackage{graphicx}

\newcommand{\HGeo}{{}_2\!F_1}
\newcommand{\Laur}{F^{(4)} _D}

\begin{document}
 \title{Wave functions for the Schwarschild black hole interior}

\author{Daniel Cartin}
\email{cartin@naps.edu}\affiliation{Naval Academy Preparatory School, 197 Elliot Street, Newport, Rhode Island 02841}

\author{Gaurav Khanna}
\email{gkhanna@UMassD.Edu}
\affiliation{Physics Department, University of Massachusetts at Dartmouth, North Dartmouth, Massachusetts 02747}

\date{\today}
\begin{abstract}

Using the Hamiltonian constraint derived by Ashtekar and Bojowald, we look for pre-classical wave functions in the Schwarzschild interior. In particular, when solving this difference equation by separation of variables, an inequality is obtained relating the Immirzi parameter $\gamma$ to the quantum ambiguity $\delta$ appearing in the model. This bound is violated when we use a natural value for $\delta$ based on loop quantum gravity together with a recent proposal for $\gamma$. We also present numerical solutions of the constraint.

\end{abstract}


\maketitle

\section{Introduction}

One of the more interesting class of solutions found in the theory of general relativity is the black hole. Not only does it fire the public imagination -- as seen by a survey of popular books of both the fiction and non-fiction variety -- but it also provides an arena to study what happens when general relativity and quantum mechanics are important. The strong gravitational fields that curve space-time enough to prevent escape from the black hole interior eventually lead to a singularity, where the classical theory becomes meaningless. Like the Big Bang singularity, a complete understanding of what occurs at this location depends on uniting classical and quantum ideas. It is thought highly unlikely that there is a complete breakdown in the equations of gravity. Instead it is hypothesized that quantum effects provide a limitation on the magnitude of space-time curvature, and perhaps even allow the possibility of discussing the region beyond the classical singularity. Thus the simplest form of black hole, the Schwarzschild space-time, is the focus of a variety of explorations into quantized gravity.

One of the more developed techniques is that of loop quantum gravity~\cite{lqg}. Unfortunately, this theory still has some outstanding issues, so a symmetry reduced version of loop quantum gravity has been developed, known as loop quantum cosmology~\cite{lqc1, lqc2}. To some extent, this allows a testbed for different incarnations of the full theory, and leads to the discovery of features that are robust under modification of the exact quantization method. Many models of cosmological interest have been studied using the wave functions of the full theory. The basic method is to start with the kinematic Hilbert space of the full theory, then reducing down to those states obeying a particular symmetry in order to quantize the Hamiltonian constraint. Research in cosmological singularities has been fruitful, and shown that there are no difficulties in resolving the singularity or indeed, in evolving the wave function through it~\cite{boj02}. Now work has begun on the singularity occurring inside the spherically symmetric black hole~\cite{ash-boj05} (see also~\cite{mod04}). This paper will show explicit solutions for the quantum constraint in the Schwarzschild interior and comment on the conditions necessary to ensure the wave function is smooth far away from the singularity, a region where quantum effects are not expected. This requirement captures the notion of {\it pre-classicality}~\cite{boj01}, ensuring the wave function has the desired physical properties when there is not yet a physical inner product applicable to this situation. The expectation is that pre-classicality picks out wave functions in the eventual physical solution space. This is due to the fact that the Wheeler-De Witt equation for semi-classical states is an asymptotic limit of the quantum constraint equation.

In Section \ref{review}, we review previous work on the interior of the Schwarzschild black hole in the context of loop quantum cosmology. The interior portion of the space-time is chosen because it is of Kantowski-Sachs type. Since these metrics are spatially homogeneous, they are very similar to other models previously considered in loop quantum cosmology. The quantum operator corresponding to the Hamiltonian constraint is a partial difference equation acting on the eigenstates of the triad, and includes a quantum ambiguity $\delta$ related to the fundamental length scale. Here, we point out the effect of using a self-adjoint constraint operator in the quantum theory. Because the wave functions will not have the same restrictions that are seen in other work (where a non-self-adjoint operator is used) the range of solutions is correspondingly greater. This choice of constraint will eventually allow the study of the physical wave functions, via the use of group averaging. In Section \ref{analytic}, we use generating function techniques to solve explicitly for wave functions. To simplify this search, we use a separation of variables method to find solutions for the constraint equation. Solving for one of the two sequences is technically demanding, so we only consider its asymptotic limit in Section \ref{asymp-a}, deferring the exact results to the Appendix. In Section \ref{numeric}, we report on numerical simulations of generic wave functions. Within the limits discussed in the paper, Gaussian wave packets can evolve under the quantum constraint, maintaining a classical trajectory until they get very close to the singularity.  Imposing the physical condition that wave functions must be suitably smooth far away from the classical singularity leads to an inequality related the quantum ambiguity $\delta$ mentioned above and the Immirzi parameter $\gamma$. Interestingly, this inequality is not satisfied for all the current proposals for $\gamma$ if a natural choice for $\delta$ is made based on the full theory. These results are discussed further in Section \ref{conclude}.

\section{The Schwarzschild interior in loop quantum cosmology}
\label{review}

A model of the Schwarzschild interior in loop quantum cosmology has been carried out by Ashtekar and Bojowald~\cite{ash-boj05}, so we will briefly review its features here. Because this portion of the space-time is spatially homogeneous, the construct is similar to other models studied in LQC. The configuration space is coordinatized by the two independent triad components $p_b$ and $p_c$; the line $p_b = 0$ gives the horizon of the black hole, while $p_c = 0$ is the singularity. The corresponding quantum operators for the triad act on their eigenstates by
\begin{equation}
\label{p-values}
{\hat p}_b  |\mu, \tau \rangle =  \frac{1}{2} \mu \gamma \ell^2 _p |\mu, \tau \rangle, \qquad
{\hat p}_c  |\mu, \tau \rangle = \gamma \tau \ell_p ^2 |\mu, \tau \rangle,
\end{equation}
where $\gamma$ is the Immirzi parameter, and $\ell_p$ is the Planck length. Thus the line $\mu = 0$ locates the event horizon while $\tau = 0$ is the classical singularity. There is a residual gauge symmetry on the phase space generated by the Gauss constraint of the model, which requires that a choice is made to fix this freedom. In previous work on loop quantum cosmology, a gauge choice such as $p_b \ge 0$ was made. This still allows passage through the classical singularity, so its resolution can be studied. However, we can also consider sequences symmetric under $\mu \to -\mu$. This does not put an artificial restriction on the extent of the quantum wave function, but picks out those that satisfy the Gauss constraint of the model. We shall see in Section \ref{numeric} that this simplifies numerical simulations of the solution.

The Hamiltonian constraint arising from loop quantum cosmology in the Schwarzschild interior acts on a triad eigenstate to give
\begin{eqnarray*}
{\hat C} ^{(\delta)} |\mu, \tau \rangle &=& (2 \gamma^3 \delta^3 \ell_p^2) ^{-1} [2(V_{\mu + \delta, \tau} - V_{\mu - \delta, \tau})(|\mu + 2\delta, \tau+2\delta \rangle - |\mu + 2\delta, \tau-2\delta \rangle - |\mu - 2\delta, \tau+2\delta \rangle + |\mu - 2\delta, \tau- 2\delta \rangle) \\
&+& (V_{\mu, \tau + \delta} - V_{\mu, \tau - \delta})(|\mu + 4\delta, \tau \rangle - 2(1 + 2\gamma^2 \delta^2) |\mu, \tau \rangle + |\mu - 4\delta, \tau \rangle)],
\end{eqnarray*}
where $V_{\mu, \tau}$ is the eigenvalue of the volume operator, given by
\[
V_{\mu, \tau} = 2 \pi \gamma^{3/2} \ell_p ^3 |\mu| \sqrt{|\tau|}.
\]
Notice that, because of the form of this constraint, the coefficients of the relation vanish at certain points. In particular, $V_{\mu + \delta, \tau} - V_{\mu - \delta, \tau} = 0$ when $\mu = 0$, and $V_{\mu, \tau + \delta} - V_{\mu, \tau - \delta} = 0$ when $\tau = 0$. This situation arises in previous work in loop quantum cosmology, such as the isotropic~\cite{boj02} and the Bianchi I and IX models~\cite{car-kha-boj04, car-kha05a, boj-dat-hos04}. If we write the wave function $\Psi$ solving the Hamiltonian constraint as a sum of eigenstates
\[
\Psi = \sum_{\mu, \tau} s_{\mu, \tau} |\mu, \tau \rangle,
\]
then we find that
\begin{eqnarray*}
&\ & 2(V_{\mu - \delta, \tau - 2\delta} - V_{\mu - 3\delta, \tau - 2\delta}) s_{\mu - 2\delta, \tau - 2\delta} + 2(V_{\mu + 3\delta, \tau + 2\delta} - V_{\mu + \delta, \tau + 2\delta}) s_{\mu + 2\delta, \tau + 2\delta} \\
&-& 2(V_{\mu - \delta, \tau + 2\delta} - V_{\mu - 3\delta, \tau + 2\delta}) s_{\mu - 2\delta, \tau + 2\delta} - 2(V_{\mu + 3\delta, \tau - 2\delta} - V_{\mu + \delta, \tau - 2\delta}) s_{\mu + 2\delta, \tau - 2\delta} \\
&+& (V_{\mu - 4\delta, \tau + \delta} - V_{\mu - 4\delta, \tau - \delta}) s_{\mu - 4\delta, \tau} + (V_{\mu + 4\delta, \tau + \delta} - V_{\mu + 4\delta, \tau - \delta}) s_{\mu + 4\delta, \tau} \\
&-& (1 + 2 \gamma^2 \delta^2) (V_{\mu, \tau + \delta} - V_{\mu, \tau - \delta}) s_{\mu, \tau} = 0.
\end{eqnarray*}

Thus $s_{0, \tau}$ and $s_{\mu, 0}$ never appear in any relation, since their coefficients always vanish in the recursion relation. As we will see below, the conditions for pre-classicality can be expressed in terms of relations between the sequence members $s_{\mu, \tau}$ for low values of $\mu, \tau$. For a simpler example, we look at the recursion relation
\[
m s_{m+1} - 2 s_{m} + m s_{m-1} = 0, \qquad m \ge 1,
\]
arising as a special case in the Bianchi I model~\cite{car-kha-boj04}. Note that there is always a scaling freedom $s_m \to c s_m$, so choosing the ratio $s_1 / s_0$ is the only free information we have in the initial values of the sequence. The sequence will oscillate without bound unless this ratio is fixed to be a certain non-zero value. Obviously this requires that both $s_0$ and $s_1$ be non-zero. When we have vanishing coefficients in the recursion relation, as with the difference operator ${\hat C} ^{(\delta)}$, a full range of smooth sequences is not available. This happens in the Bianchi I case, where $s_0 = 0$ for exactly this reason. When we consider the self-adjoint version of the constraint, however, this situation does not arise and there are no vanishing coefficients of the difference operator. So the boundary value $s_0$ is included in the partial difference equation coming from the Hamiltonian constraint and can be non-zero.

This inclusion of the boundary values allows a much greater range of pre-classical sequences, as we will see later. However, in addition to this practical utility, the use of a self-adjoint version also helps in making contact with the physical Hilbert space of solutions. This is through the use of group averaging techniques~\cite{grp-avg}, already used in a few applications to loop quantum cosmology~\cite{grp-avg-lqc}. Group averaging is a method to explicitly construct the physical solutions of a quantized and constrainted system. One picks out those functions solving the constraint by averaging over the one-parameter group generated by the constraint; this cannot be accomplished unless the constraint is self-adjoint.

From this point on, we work with the symmetric operator ${\hat {\cal H}} = \frac{1}{2} ({\hat C} ^{(\delta)} + {\hat C}^{(\delta) \dag})$. We write the constraint equation ${\hat {\cal H}} |\mu, \tau \rangle = 0$ in terms of new parameters $m, n$, where $\mu = 2m \delta$ and $\tau = 2n\delta$. This results in the following recursion relation for all $m, n$:
\begin{eqnarray}
\label{full-diff}
\nonumber
&\ & \biggl[ \biggl(|m - \frac{1}{2}| - |m - \frac{3}{2}| \biggr) \sqrt{|n - 1|} + \biggl(|m + \frac{1}{2}| - |m - \frac{1}{2}| \biggr) \sqrt{|n|} \biggr] s_{m - 1, n - 1} \\
\nonumber
&-& \biggl[ \biggl(|m - \frac{1}{2}| - |m - \frac{3}{2}| \biggr) \sqrt{|n + 1|} + \biggl(|m + \frac{1}{2}| - |m - \frac{1}{2}| \biggr) \sqrt{|n|} \biggr] s_{m - 1, n + 1} \\
\nonumber
&-& \biggl[ \biggl(|m +  \frac{3}{2}| - |m + \frac{1}{2}| \biggr) \sqrt{|n - 1|} + \biggl(|m + \frac{1}{2}| - |m - \frac{1}{2}| \biggr) \sqrt{|n|} \biggr] s_{m + 1, n - 1} \\ 
&+& \biggl[ \biggl(|m +  \frac{3}{2}| - |m + \frac{1}{2}| \biggr) \sqrt{|n + 1|} + \biggl(|m + \frac{1}{2}| - |m - \frac{1}{2}| \biggr) \sqrt{|n|} \biggr] s_{m + 1, n + 1} \\ 
\nonumber
&+& \frac{1}{2} \biggl[ \biggl( |m - 2| + |m| \biggr) \biggl( \sqrt{| n + \frac{1}{2}|} - \sqrt{|n - \frac{1}{2}|} \biggr) \biggr] s_{m-2, n} \\
\nonumber
&-& (1 + 2 \gamma^2 \delta^2) |m| \biggl( \sqrt{|n + \frac{1}{2}|} - \sqrt{|n - \frac{1}{2}|} \biggr) s_{m, n} \\
\nonumber
&+& \frac{1}{2} \biggl[ \biggl( |m + 2| + |m| \biggr) \biggl( \sqrt{| n + \frac{1}{2}|} - \sqrt{|n - \frac{1}{2}|} \biggr) \biggr] s_{m+2, n} = 0.
\end{eqnarray}
Because the original parameters $\mu, \tau$ can take any real value, there is no loss of generality by using the scaled variables $m, n$. However, notice that the quantum ambiguity $\delta$ and the Immirzi parameter $\gamma$ drop out almost everywhere when $m, n$ are used, except for in the coefficient for $s_{m, n}$. This will be the only place where physical input will affect the solution. In addition, we note that the constraint greatly simplifies when $|m| > 3/2$ a fact we will use in the next section. Now we turn to the task of solving this difference equation.

\section{Analytic solutions}
\label{analytic}

\subsection{Generating function techniques}

In general, it is difficult to find solutions to multi-parameter recursion relations where the sequence is smooth for large values of the parameters. A generic feature are sign flips every time the sequence parameter is increased by one step. One effective means to study the behavior of these sequences is to employ a separation of variables technique, analogous to the procedure used for partial differential equations. The partial difference equation is then reduced to several one-parameter recursion relations, which are much easier to deal with. Examining the self-adjoint constraint for the Schwarzschild interior, we see that when we assume that $m \ge 3/2$ (i.e. consider the "bulk" of the space-time), the relation is separable in this manner. Denoting $s_{m, n} = \alpha_m \beta_n$, we get
\begin{subequations}
\begin{eqnarray}
\label{a-true-eqn}
(m + 1) \alpha_{m + 2} - (1 + 2 \gamma^2 \delta^2) m \alpha_m + (m - 1)\alpha_{m - 2} &=& \lambda (\alpha_{m + 1} - \alpha_{m - 1}) \\
\label{b-true-eqn}
(\sqrt{|n + 1|} + \sqrt{|n|}) \beta_{n + 1} - (\sqrt{|n - 1|} + \sqrt{|n|}) \beta_{n - 1} &=& - \lambda \biggl( \sqrt{|n + \frac{1}{2}|} - \sqrt{|n - \frac{1}{2}|}\biggr) \beta_n,
\end{eqnarray}
\end{subequations}
where $\lambda$ is a separation parameter. We can obtain solutions to these two recursion relations by using generating function methods~\cite{car-kha-boj04}, which are reviewed later in this section. However, to use these techniques, the coefficients of the sequence values $\beta_n$ must be polynomial. We factor out $\sqrt{|n|}$ from the equation, and use a Taylor series expansion of the resulting coefficients, keeping only up to order $O(1/n)$ and considering only positive $n$ for the moment\footnote{Here we make a comment about the accuracy of using this approximate relation for $\beta_n$. When we look at the solution $\beta_n$ at large $n$, the errors between the order $1/n$ relation and the completely accurate equation will be very small, and limited to the regime where $n \sim O(1)$. Thus the difference between any solution we find for our relations here versus the relation for all $m, n$ is noticeable only close to the singularity $n=0$. Similar reasoning lets us ignore the difference between the full recursion relation and the "bulk" relation, where $m \ge 2$ allows us to simplify the absolute value signs for $m$.}. This gives us
\begin{equation}
\label{step2}
\biggl(2 + \frac{1}{2n} \biggr) \beta_{n+1} - \biggl(2 - \frac{1}{2n} \biggr) \beta_{n-1}  = - \frac{\lambda}{2n} \beta_n, \qquad n > 0.
\end{equation}
Obviously this does not work when $n=0$; when we look back at the original separated relation (\ref{b-true-eqn}) for $\beta_n$ with $n=0$, we get the simple relation that $\beta_1 = \beta_{-1}$. Finally, we can multiply our approximate relation (\ref{step2}) for $\beta_n$ by $2n$ in order to give coefficients polynomial in $n$; because $n > 0$, this does not affect the resulting solutions. Thus, the two relations we will solve for separable sequences are the relation (\ref{a-true-eqn}) for $\alpha_m$, and
\begin{equation}
\label{b-eqn}
(1 + 4n) \beta_{n+1} + (1 - 4n) \beta_{n-1} = -\lambda \beta_n
\end{equation}
for $\beta_n$. Because the recursion relation for the $\beta_n$ sequence is simpler -- a second order relation as opposed to the fourth order difference equation for $\alpha_m$ -- we will search for its solutions first.

Solutions for a recursion relation in general will have undesirable physical properties. Generically these sequences will have oscillatory behavior -- adjacent values will have the opposite sign, and the magnitude of the sequence values may increase without bound as the sequence parameter increases. At this time, there is limited work on finding a physical inner product in loop quantum cosmology, so there is no way to pick out such unphysical states. This leads to the notion of pre-classicality~\cite{boj01}, where we put in place criteria that pick out wave functions with desirable properties. The expectation is that pre-classicality will pick out states that appear in the physical Hilbert space. This is based on the fact that the quantum Hamiltonian constraint is a discretized version of the Wheeler-De Witt equation~\cite{lqc1}; in the limit of vanishing step size, solutions of the difference equation will go to those of the semi-classical differential equation. Thus, those sequences that have a smooth limit will match wave functions solving the Wheeler-De Witt equation.

Because we are looking for sequences that represent the wave function of a space-time, which becomes classical for large volumes (i.e. far away from any singular points), we must have a way of restricting the wave functions to those that act semi-classically in the appropriate regime. With this in mind, the technique of using generating functions to solve difference equations has been developed in the context of loop quantum cosmology~\cite{car-kha-boj04}; for a review of these methods in a generic context, see Wilf~\cite{wil93}. Our goal will be to find a generating function $B(y)$, whose Taylor series\footnote{Notice that we are focussing solely on integer values of the parameter $k$, despite the fact that $\mu$ and $\tau$ (and hence $m, n$) can take any real value. It has been shown elsewhere~\cite{car-kha05b} that using the sequence solution for integer values can be extended to all real numbers.} gives the sequence $\beta_n$, i.e.
\[
B(y) = \sum_{k=0} ^\infty \beta_k y^k.
\]
Operations on the sequence $\beta_n$ can be mapped over to those on the function $B(y)$. For example, suppose we want to find the generating function for the sequence ${\tilde \beta}_n = \beta_{n+1}$; thus, we find that
\[
\sum_{k=0} ^\infty {\tilde \beta}_n y^n = \sum_{k=0} ^\infty \beta_{n+1} y^n = \sum_{k=1} ^\infty \beta_n y^{n-1} = y^{-1} \biggl( \sum_{k=0} ^\infty \beta_n y^n - \beta_0 \biggr).
\]
Thus the shifting operation $\beta_n \to \beta_{n+1}$ is equivalent to the operation $B(y) \to [B(y) - \beta_0]/y$ on the generating function. Similarly, if we want a multiplication operator ${\tilde \beta}_n = n\beta_n$, we get
\[
\sum_{k=0} ^\infty {\tilde \beta}_n y^n = \sum_{k=0} ^\infty n \beta_n y^n = y \frac{d}{dy} \sum_{k=0} ^\infty \beta_n y^n.
\]
So multiplication $\beta_n \to n\beta_n$ corresponds to using the Euler operator $y d/dy$ acting on $B(y)$. This is the reason we used the earlier approximation for the recursion relation; to get a differential equation for $B(y)$ that is relatively easy to solve, we must stick to using operators that are polynomial in both the variable $y$ and the Euler operator.

Up to this point, the Taylor series is merely a formal device to get a solution to the generating function. However, we have also a physical requirement that the sequence is smooth for large values of the parameter, corresponding to classical space-times. This will correspond to limitations on the singularities of the function $B(y)$, which can be seen in the following examples. Suppose we have a simple pole at $y=a$, where $0 < |a| \le 1$; the Taylor series of the monomial $(1 - y/a)^{-1}$ alternates in sign if $a<0$, and increases without bound when $|a| \le 1$, since
\begin{equation}
\label{seq-ex}
\biggl(1 - \frac{y}{a} \biggr)^{-1} = \sum_{k=0} ^\infty {-1 \choose k} \biggl( -\frac{y}{a} \biggr)^k = \sum_{k=0} ^\infty \biggl( \frac{y}{a} \biggr)^k
\end{equation}
Higher order poles makes the problem worse, since the magnitude of the coefficients of $y^n$ appearing in the Taylor series would increase accordingly. This gives an equivalence between singularities of the generating function $B(y)$ and the asymptotic behavior of the sequence $\beta_n$. A singularity at $y=1$, for example, shows this directly, since
\[
(1 - y) B(y) |_{y=1} = \beta_0 + \sum_{n=0} ^\infty (\beta_{n+1} - \beta_n) (1)^{n+1} = \lim_{n \to \infty} \beta_n.
\]
When $B(y)$ has a pole at $y=1$ that is of order one or less, then the sequence $\beta_n$ approaches a finite value; if the pole has an order greater than one, the sequence is unbounded. Similarly, we can look for poles at $y=-1$ to determine whether the sequence is oscillatory, flipping signs as the parameter $n$ is incremented by one. If the generating function includes a monomial $(1 + y)^{-p}$ having an order $p$ close to one, oscillations may occur that eventually dampen out. In general, requiring that $B(y)$ is finite whenever $-1 \le y < 0$ will ensure the corresponding sequence $\beta_n$ has the appropriate semi-classical behavior. If we wish to have only bounded sequences, then we can add the further condition that $B(y)$ is finite for $0 < y \le 1$. Thus our criteria for the pre-classicality of a solution to the Hamiltonian constraint will be phrased in terms of the singularity structure of its associated generating function.

We return to a simple example to illustrate what is happening, the previously mentioned relation
\[
m s_{m+1} - 2 s_{m} + m s_{m-1} = 0,
\]
which arises from separation of the Hamiltonian constraint in Bianchi I LRS. The generating function for this relation is of the form
\[
F(x) = \frac{a_0 - (2a_0 + a_1)x - (4a_0 + 2a_1) \ln(1-x)}{(1+x)^2} = \frac{C(x)}{(1+x)^2},
\]
where the function $C(x)$ depends on two constants $a_0$ and $a_1$ (these are the first two values of the sequence). Thus we have to worry about a second order pole at $x=-1$, and we must pick a relation between $a_0$ and $a_1$ to ensure the function $F(x)$ is finite at this point. The reason we can do this is obvious when we Taylor expand $C(x)$ around the point $x=-1$:
\[
C(x) = c_0 + c_2 (1 + x)^2 + \cdots.
\]
Because $C(x)$ has no term that is first order in $(1+x)$, then we can make $c_0 = 0$ by the appropriate choice of constants $a_0$ and $a_1$. Once this is done, the generating function $F(x)$ is finite at $x=-1$, and the associated sequence will not have growing oscillatory behavior. If there was a non-zero term $c_1 (1+x)$ in the expansion, the only solution would have been $C(x) = 0$, and the sequence would have zero for all its values. Once the pole at $x=-1$ is taken care of, there is only the pole at $x=1$ due to the logarithm function. However, since $(1-x) \ln(1-x) = 0$ at $x=1$, then $\lim_{m \to \infty} a_m = 0$ implies the sequence asymptotically approaches zero.

\subsection{$\beta_n$ sequence}

In this manner we shall find a differential equation for the generating function, with the appropriate conditions on the order and location of any singularities. We make the shift $B(y) = \beta_0 + y F(y)$ for convenience; from this we get the condition $F(0) = \beta_1$. In terms of $F(y)$, the equation (\ref{b-eqn}) for $\beta_n$ corresponds to
\begin{equation}
4y(1 - y^2) \frac{\partial F(y)}{\partial y} + F(y) (1 + \lambda y - 7y^2) - 3 \beta_0 y - \beta_1 = 0
\end{equation}
Solving this equation for $F(y)$ gives
\begin{eqnarray}
\nonumber
F(y) &=&  - \frac{1}{4} y^{-1/4} (y - 1)^{-3/4 + \lambda/8} (y + 1)^{-3/4 - \lambda/8} \biggl[ \int^y _0 (3 \beta_0 z + \beta_1) z^{-3/4} (z - 1)^{-1/4 - \lambda/8} (z + 1)^{-1/4 + \lambda/8} dz \biggr] \\
\label{b-ODE}
&=& (1 - y)^{-3/4 + \lambda/8} (1 + y)^{-3/4 - \lambda/8} \biggl[ \beta_1 F_1 \biggl(\frac{1}{4}; \frac{2 + \lambda}{8}, \frac{2 - \lambda}{8}; \frac{5}{4}; y, -y \biggr) \\
\nonumber
&+& \frac{3 \beta_0 y}{5} F_1 \biggl(\frac{5}{4}; \frac{2 + \lambda}{8}, \frac{2 - \lambda}{8}; \frac{9}{4}; y, -y \biggr)\biggr].
\end{eqnarray}
To ensure that $F(0) = \beta_1$, the integration constant obtained when integrating is set to zero. $F_1$ is the Appell hypergeometric function of two variables~\cite{ext76}, defined as
\[
F_1 (a; b_1, b_2; c; z_1, z_2) = \sum_{j = 0} ^{\infty} \sum_{k = 0} ^{\infty} \frac{ (a)_{j+k} (b_1)_j (b_2)_k}{(c)_{j+k} j! k!} z_1 ^j z_2 ^k
\]
where $(x)_n = \Gamma(x + n)/\Gamma(x)$ is the Pochhammer symbol (or rising factorial), involving the gamma functions $\Gamma(x)$. Although an analytic continuation of this function can be defined for any values of the variables $z_1$ and $z_2$, as a {\it series}, $F_1$ converges only when $|z_1| < 1$ and $|z_2| < 1$. All of the singular behavior of the function $B(y)$ generating the values of $\beta_n$ is contained in the function $F(y)$. In this way, we can find all the properties of the sequences solving the relation (\ref{b-eqn}) by studying the function $F(y)$ solving a differential equation. Notice that when we solve for the function $F(y)$, we find all values of the sequence simultaneously. Once the generating function is obtained, all values of the sequences can be read off the Taylor series expansion. This compares to using the recursion relation to find $\beta_n$, which would require first having $\beta_{n-1}$ and $\beta_{n-2}$; the relation is inherently an evolution equation with a sense of "time". The equation for $F(y)$ is parametrized by the initial values $\beta_0$ and $\beta_1$, but it is not required we use these. Now that there is no restriction on $\beta_0$, as there would be in the non-self-adjoint case, we could use $\beta_{N-1}$ and $\beta_N$ as our parameters, where $N$ is a large positive integer (see~\cite{wil93} for more discussion).

Now we locate the singularities of the function $F(y)$; because of the polynomial appearing in front of the derivative in the differential equation, these are at $y = \pm 1$. Our only true degree of freedom is the ratio $\beta_1 / \beta_0$, since the sequence can be scaled by an arbitrary value, so we focus on keeping the generating function finite at $y=-1$ to avoid growing oscillations. When evaluating $F(-1)$, there are the obvious poles due to the monomials in the function. For $\lambda < 6$, there is a singularity at $y=-1$; similarly, there is another at $y=1$ when $\lambda > -6$. However, the Appell functions also have their own singularities at $y = \pm 1$; to evaluate these, we use the fact~\cite{ext76} that in the limit $z_1 \to 1$
\begin{eqnarray}
\nonumber
\lim_{z_1 \to 1} F_1 (a; b_1, b_2; c; z_1, z_2) &=& \frac{\Gamma(c) \Gamma(a + b_1 - c)}{\Gamma(a) \Gamma(b_1)} (1 - z_1)^{c-a-b_1} (1 - z_2)^{-b_2} (1 + O(z_1 - 1)) \\
\label{expand}
&+& \frac{\Gamma(c) \Gamma(c - a - b_1)}{\Gamma(c - a) \Gamma(c - b_1)} \HGeo (a, b_2; c - b_1; z_2) (1 + O(z_1 - 1))
\end{eqnarray}
where $\Gamma(n)$ is the gamma function and $\HGeo$ the Gaussian hypergeometric function. Recall that $\HGeo (a, b; c; x)$ is convergent only when the real part of $(c - a - b) > 0$~\cite{abr-ste}. Using the particular Appell functions appearing in our generating function, and the exchange symmetry
\[
F_1 (a; b_1, b_2; c; z_1, z_2) = F_1 (a; b_2, b_1; c; z_2, z_1),
\]
we find they are singular at $y=1$ when $\lambda>6$, and at $y=-1$ when $\lambda<-6$. Thus, these functions have exactly the opposite behavior as the monomials. However, notice what happens as we vary $\lambda$, for the $y=-1$ pole in particular. When $\lambda < -6$, the monomial $(1+y)^{-3/4 - \lambda/8}$ goes to zero as $y \to -1$. As we can see from the behavior of the Appell function as $z_1 \to 1$, and the exchange symmetry, the two Appell functions diverge as $(1+y)^{3/4+\lambda/8}$. Thus, the two conspire so that the generating function as a whole remains finite at $y = -1$. Because there is no singular behavior for $\lambda < -6$, the associated sequence $\beta_n$ will not be oscillatory for {\it any} choice of values for $\beta_0$ and $\beta_1$. However this does not occur when $\lambda > -6$. In that case, the monomial now diverges; the Appell functions are finite but non-zero, so there is no cancellation and $F(-1)$ is divergent. Therefore the ratio $\beta_1/\beta_0$ cannot be arbitrary, but must be chosen by requiring
\begin{equation}
\label{b-cond}
\beta_1 F_1 \biggl(\frac{1}{4}; \frac{2 + \lambda}{8}, \frac{2 - \lambda}{8}; \frac{5}{4}; -1, 1 \biggr) - \frac{3 \beta_0}{5} F_1 \biggl(\frac{5}{4}; \frac{2 + \lambda}{8}, \frac{2 - \lambda}{8}; \frac{9}{4}; -1, 1 \biggr) = 0.
\end{equation}
This can be simplified by using the gamma functions of the second term in the expansion (\ref{expand}) around $z_1 = 1$ after exchanging $z_1$ and $z_2$, giving
\[
\frac{\beta_1}{\beta_0} = \biggl( \frac{6}{\lambda + 8} \biggr) \frac{\HGeo(\frac{5}{4}; \frac{1}{4} + \frac{\lambda}{8}, 2 + \frac{\lambda}{8}; -1)}{\HGeo(\frac{1}{4}; \frac{1}{4} + \frac{\lambda}{8}, 1 + \frac{\lambda}{8}; -1)}, \qquad \lambda > -6.
\]

So, we find that the ratio $\beta_1/\beta_0$ is completely free when $\lambda < -6$, but must be fixed otherwise to avoid oscillations far from the singularity $n=0$. We can expand the function $B(y) = \beta_0 + y F(y)$ in a Taylor series, to read off the values of the sequence $\beta_n$ for any real $n$, as done in previous work~\cite{car-kha05b}. However, in this case the function is not easily written in a compact form. Also, by examining the pole at $y=1$ in a manner similar to the above, we can see whether the resulting sequence $\beta_n$ is bounded or not. This gives us that the divergence of the Appell functions is cancelled out by the monomial going to zero when  $\lambda > 2$, so $F(1)$ is finite in this regime, and the sequence goes to zero asymptotically. Otherwise, the sequence will increase without bound. Finally, we comment here that having a non-zero value $\beta_0$ is crucial for obtaining pre-classical solutions for all values of the separation parameter $\lambda$. If $\beta_0 = 0$ (as would be the case if a non-self-adjoint constraint were used), then the only sequence meeting the pre-classicality condition (\ref{b-cond}) would be the trivial one $\beta_n = 0$. This would severely restrict the space of solutions, since physical wave functions would require $\lambda \le -6$. 

\subsection{Asymptotic limit of the $\alpha_m$ sequence}
\label{asymp-a}

When we treated the case of the $\beta_n$, the physical parameters did not enter anywhere in the analysis. Thus, the results there are independent of the Immirzi parameter $\gamma$ and the ambiguity $\delta$ arising in the quantization. Because of the reparametrization we chose -- going from the original triad eigenvalues $\mu$ and $\tau$ over to the new variables $m, n$ -- the only place these appear is in the recursion relation for the separable sequence $\alpha_m$ in the combination $\gamma \delta$. As we will see, this will tie the existence of solutions with the proper semi-classical behavior to the values of $\gamma$ and $\delta$, and put a limit on their product.

Because the study of the $\alpha_m$ sequence is more involved, we look here at the asymptotic behavior of the sequence, and reserve the full details to the Appendix. This will give us a roadmap to the results to be obtained for all values of the triad eigenvalue $m$ using generating function techniques. By looking at the limiting cases of the sequences for large $m$, we will reproduce the division of behavior into unbounded and decaying sequences that occurs in the full solutions. To start with, we think of the recursion relation (\ref{a-true-eqn}) for $\alpha_m$ as an equation for a function $a(m)$. We use an expansion of $a(m)$ in terms of a step size $h$, for example,
\[
\alpha_{m+2} \to a(m + 2h) = a(m) + 2 \frac{d a(m)}{dm} h + 2 \frac{d^2 a(m)}{d^2 m} h^2 + O(h^3).
\]
By doing the same for the rest of the relation, we obtain a differential equation for $a(m)$ to various orders of $h$; in the following, we go up to second order in $h$, and set $h=1$. This results in
\begin{equation}
4m \frac{d^2 a(m)}{d m^2} + (4 - 2 \lambda) \frac{d a(m)}{d m} + (2 - \kappa) a(m) = 0,
\end{equation}
where we define $\kappa = 1 + 2 \gamma^2 \delta^2$. Since the equation simplifies if we choose $\kappa = 2$, let us solve it first for this case, giving
\[
a(m) = \alpha_0 + [\alpha_1 - \alpha_0]m^{\lambda/2} + O(m^{\lambda/2 - 1}),
\]
with the constants set using the first two values of the sequence $\alpha_m$. Thus, already we can see what happens for large values of the triad parameter $m$ when $\kappa = 2$ -- if $\lambda$ is positive, the sequence will increase without limit, while if $\lambda$ is negative, it will decay to zero.

When $\kappa \ne 2$, the solution to the equation is
\[
a(m) = m^{\lambda/4} \biggl[ C_1 J_{\lambda/4} \biggl(\frac{\sqrt{2 - \kappa}}{2} m \biggr) + C_2 Y_{\lambda/4} \biggl(\frac{\sqrt{2 - \kappa}}{2} m \biggr)\biggr] + O(m^{\lambda/4 - 1}),
\]
where $J_\nu (x)$ and $Y_\nu (x)$ are Bessel functions of the first and second kind, respectively, and $C_1, C_2$ are two constants of integration. Note that, when $\kappa > 2$, the arguments of these functions will become imaginary, so the solution of the equation will feature the modified Bessel functions $I_\nu (x)$ and $K_\nu (x)$. We examine the large $m$ behavior of $\kappa < 2$ first. In this case, as $x \to \infty$,
\[
x^{\lambda/4} J_{\lambda/4} \biggl(\frac{\sqrt{2 - \kappa}}{2} x \biggr) \sim x^{-1/2 + \lambda/4} \cos \biggl(\frac{\sqrt{2 - \kappa}}{2} x - \frac{\lambda \pi}{8} - \frac{\pi}{4} \biggr),
\]
and similarly for $Y_\nu (x)$. Thus, we expect to find sequences that oscillate over a wavelength of $4 \pi / \sqrt{2 - \kappa}$, as it either decays $(\lambda < 2)$ or increases in amplitude $(\lambda > 2)$. Notice that our criterion for pre-classicality is to avoid the sequence changing signs as the parameter $m$ increases by one step; a gentler sinusoidal oscillation is no grounds for discarding a solution since this often seen in quantum mechanics. On the other hand, when $\kappa > 2$, the Bessel function $I_\nu (x)$ grows without bound for large $x$, since
\begin{equation}
\label{big-exp}
x^{\lambda/4} I_{\lambda/4} \biggl(\frac{\sqrt{\kappa - 2}}{2} x \biggr) \sim x^{\lambda/4 - 1/2} \exp \biggl( \frac{\sqrt{\kappa - 2}}{2} x \biggr),
\end{equation}
while the other function $K_\nu (x)$ exponentially decays. Since the exponential $e^x$ increases faster than any power of $x$, then the sequence will be unbounded for $\kappa > 2$, regardless of the value of $\lambda$. Thus, what we expect when we find complete solutions is that the $\alpha_m$ sequence will change its behavior as $\kappa$ crosses the critical value $\kappa = 2$, from a slowly oscillating function to an unbounded one.

\section{Numerical solutions}
\label{numeric}

Analyzing the separable solutions has enabled us to understand the general properties of pre-classical solutions of the Hamiltonian constraint. However, the structure of the self-adjoint constraint derived for this model is very beneficial in using numerical techniques. We now proceed to find solutions in this manner, using the parameter $n$ (or $\tau$) as a time parameter. An arbitrary wave packet is chosen at some relatively large distance away from both the classical singularity and the horizon. Specifically we pick a profile $s_{m, N}$ for a fixed large value $N$. This is done so that we can use the semi-classical approximation of the constraint, i.e. the Wheeler-de Witt equation given in~\cite{ash-boj05}, to find the derivative in $n$ of the packet as it is evolved towards the singularity.

Here we note several points that make the numerical calculation much easier. First, since we are using a self-adjoint constraint, there are no limitations on the values of the solution at the classical singularity. In previously considered cases, such as Bianchi class A models~\cite{boj-dat-van}, those coefficients $s_{n_1, n_2, n_3}$ of the wave function corresponding to the zero volume basis elements drop out of the recursion relation, because of various factors of the volume eigenvalue appearing in the difference equation. For convenience in solving the equations in those models, the $s_{n_1, n_2, n_3}$ include a factor of the volume, which means that $s_{n_1, n_2, 0} = 0$ and similarly for the other boundaries. Meeting this requirement would mean the profile chosen far away from the singularity would have to be exactly right or else it would "miss" the correct boundary condition. This does not occur in the self-adjoint case -- the relevant coefficients of the difference equation never vanish. Because there is no longer a restriction on the values at the singularity of the Schwarzschild interior, it is much easier to evolve arbitrary wave profiles.

In a similar vein, the range of triad eigenvalues used to delimit the configuration space in the Schwarzschild model also results in an easier problem. When solving the constraints, there are some residual symmetries between the coefficients of the wave function~\cite{boj03}. The usual choice in previous loop quantum cosmology work has been to truncate the configuration space, and allowing only non-negative values of some of the coordinates. In the diagonal Bianchi class A models, for the eigenvalues of the triad operators, it is assumed that $n_1, n_2 \ge 0$ (the third eigenvalue is unrestricted). Instead, here the wave functions are unaffected by the gauge transformation $m \to -m$, coming from the Gauss constraint~\cite{ash-boj05}. Putting this together, we can fix the boundaries of our numerical simulation to be $m = \pm M$, for large M. This leads to the second simplification in numerically solving the equations. In the original work of Ashtekar and Bojowald, there is a discussion about the boundary conditions to be imposed on the wave functions. In particular, they make an argument for choosing $s_{m, n} \to 0$ as $m \to \infty$. The physical rationale for this choice is as follows. To make contact with classical general relativity, we want the ability to construct wave packets that represent a semi-classical wave function, peaked around the classical trajectories in the phase space. These are of the form~\cite{ash-boj05}
\begin{equation}
\label{class-BH}
p_b (t) = p_b ^{(0)} \sqrt{t (2m - t)}, \qquad \qquad p_c (t) = \pm t^2,
\end{equation}
where $m$ is the mass of the black hole and $t$ an affine parameter, while $p_b ^{(0)}$ is a scaling factor that can be absorbed into the radial coordinate of the metric. We should not have solutions increasing monotonically for large $p_b$, since $p_b (t)$ arcs back to zero as $t \to 2m$.
Thus, as long as the maximum grid size $M$ is large enough to avoid sizeable errors in solving the difference equation, we can set $s_{\pm M, n} = 0$. This gives all the information necessary to find a full solution for the Hamiltonian constraint. Here again, we see the problem of using $\kappa > 2$; this boundary condition would be impossible to enforce with only unbounded sequences with $\lambda$-dependent slopes. It can be done in the $\kappa < 2$ case, where all the sequences have the same asymptotic period of oscillation.

\begin{figure}[hbt]
	\includegraphics[width=0.75\textwidth]{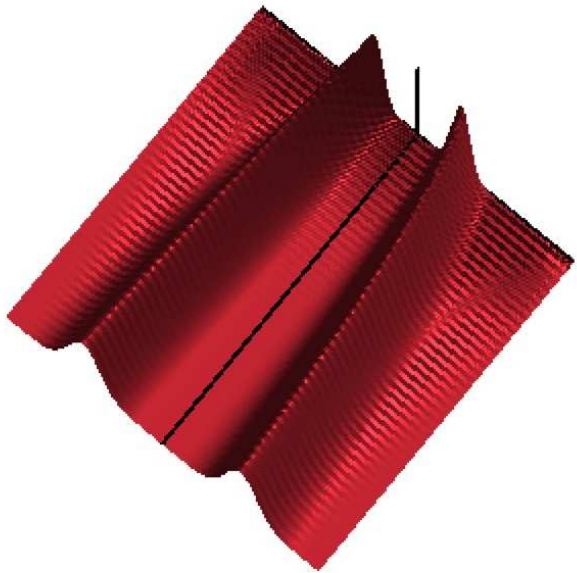}
	\caption{\label{horns}A numerical simulation of an $m$-symmetric Gaussian wave function as it approaches the singularity of the black hole. The black line in the middle represents the horizon and the classical singularity is to the upper right.}
\end{figure}

\begin{figure}[hbt]
	\includegraphics[width=0.5\textwidth]{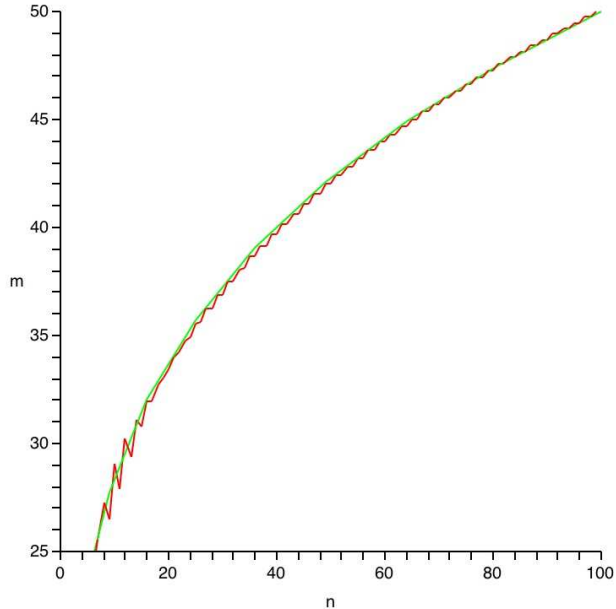}
	\caption{\label{compare}A comparison between the classical black hole and the quantum wave function. The green line indicates the classical trajectory given by the relations (\ref{class-BH}), while the red line is the averaged middle point of a Gaussian wave packet numerically evolved using the quantum constraint (\ref{full-diff}).}
\end{figure}

When all these considerations are taken into account, one can obtain a numerical solution to the full Hamiltonian constraint, shown in Figure \ref{horns}. In order to compare this to the classical case, we use the relations (\ref{class-BH}) for the phase space trajectory, solve these for a relationship between the two momenta $p_b$ and $p_c$, and compare this to the average value of the triad eigenvalues, $\mu$ and $\tau$ (or equivalently, $m$ and $n$), recalling the relations (\ref{p-values}) between the operator equivalents of $p_b$ and $p_c$, and their eigenvalues $\mu$ and $\tau$. This comparison is done in Figure \ref{compare} for a particular numerical wave function. As can be seen, the classical solution is a good approximation to the quantum wave function even very close to the classical singularity. Both the analytic and numerical sides have shown us a rich variety of wave functions that solve the Hamiltonian constraint for the Schwarzschild interior. From the form of the full constraint (\ref{full-diff}), we can see that the wave functions will be symmetric on both sides of the classical singularity. This has implications for important issues in black hole physics, in particular, as seen in a recent proposal for a paradigm of information loss suggest by loop quantum cosmology results~\cite{ash-bojBH}.

\section{Discussion}
\label{conclude}

In the preceding work we used generating function techniques to analyze the behavior of solutions to the quantum Hamiltonian constraint in the Schwarzschild interior. This was done by using a separation of variables method, and studying the resulting one-parameter sequences. The sequence $\beta_n$ does not depend on any of the physical parameters $\gamma$ and $\delta$ of the theory, so has the same form regardless of what particular case we are dealing with. On the contrary, the combination $\kappa = 1 + \gamma^2 \delta^2$ serves as a determinant for the asymptotic behavior (far away from the event horizon) of the $\alpha_m$ sequence. The goal is to use these separable solutions to assemble a semi-classical wave packet, which is peaked around the classical trajectory, so that for large $m, n$, the solution acts according to the equations of general relativity. Because these trajectories in phase space are entirely within a compact region, we expect that the wave function will act in a similar manner. With this in mind, whether or not the separable sequences go to zero as $m, n \to \infty$ is important. Since only $\alpha_m$ changes its properties with $\kappa$, we focus on it for the remainder.

If $\kappa > 2$, then aside from two isolated solutions when $\lambda = \pm \lambda_0 = \pm \sqrt{\kappa + 2}$ (see the Appendix), all solutions for $\alpha_m$ are unbounded sequences with an asymptotically exponential profile. This means that it is impossible to construct a generic wave packet to represent the black hole wave function for these values of $\kappa$. There is no way to match a linear combination of these separable sequences to sum up to a constant value at large $m$ because different values of $\lambda$ will give sequences with different slopes in the limit, as given by the asymptotic function (\ref{big-exp}) for the $\kappa > 2$ sequence. On the other hand, when $\kappa \le 2$, there is a pre-classical solution for any value of the initial data and the separation constant $\lambda$, so arbitrary wave functions are easily constructed. Hence we have found there is a relation between the values of the Immirzi parameter $\gamma$ and the ambiguity $\delta$ arising in the quantization of the Schwarzschild interior; generic solutions will only exist when
\[
\gamma \delta \le 1/\sqrt{2}.
\]
From the full theory of loop quantum gravity, using the smallest eigenvalue of the area operator and comparing it to the Schwarzschild case, we find that $\delta = 2 \sqrt{3}$~\cite{ash-boj05}. The inequality above then gives
\begin{equation}
\label{bound}
\gamma_{max} = \frac{1}{2\sqrt{6}} = 0.204124 \dots
\end{equation}
Thus, by the imposition of a boundary condition coming from the need to link up loop quantum cosmology with semi-classical general relativity, an explicit bound is placed on the Immirzi parameter. Although this is a tentative result -- in that wave functions in the interior only are considered without reference to the outside -- it is interesting to see that $\gamma$, a parameter somewhat analogous to the $\theta$ parameter in non-Abelian field theories, can be bounded by physical considerations.

The value of the Immirzi parameter is typically obtained by making contact with the Bekenstein-Hawking equation for the entropy of a black hole, that is $S_{BH} = A/4$, where $A$ is the area of the black hole event horizon. Since entropy is obtained by the logarithm of the number of states contributing to a macroscopic state, the question is how to count these states. In loop quantum gravity, the area of a surface depends on the number of spin network edges that puncture the surface. Each edge is labelled by a spin representation $\{j_k\} \in Z/2$, for $k = 1, \cdots, N$, giving a surface area of
\begin{equation}
\label{A-op}
A = 8 \pi \gamma \ell_P ^2 \sum_{k=1} ^N \sqrt{ |j_k| (|j_k| + 1)}.
\end{equation}
Originally, the assumption was that for large surface areas, the lowest spin values would predominate the sum. However, this was shown to be incorrect by Meissner~\cite{mei04}. In the limit of large area, counting all possible combinations of the spin labels give the equation
\begin{equation}
\label{mei-eqn}
1 = \sum_{k=1} ^\infty e^{- 2 \pi \gamma_M \sqrt{k (k+2)}}
\end{equation}
for the Immirzi parameter $\gamma_M$. Solving this numerically gives $\gamma_M = 0.237532 \dots$, which violates the bound (\ref{bound}) obtained here.

With this in mind, it is worthwhile to note some of the other proposals for the Immirzi parameter, where changes in the method of counting states is considered. For example, when it was assumed that the lowest order spin labels comprise the vast majority of entries in the sum, a correspondence between the area operator spectrum (\ref{A-op}) and the classical quasi-normal oscillation modes of a black hole was observed~\cite{qnm}. The argument is the following. In the limit of large damping, the real part of the quasi-normal mode frequencies becomes $\omega_{QNM} = \ln 3 / 8 \pi$. On the other hand, adding or subtracting a single puncture gives a change in area of
\begin{equation}
\label{low-spin}
\Delta A = 8 \pi \gamma \ell_p ^2 \sqrt{ |j_{min}| (|j_{min}| + 1)}.
\end{equation}
Equating these two results using Bohr's correspondence principle, one can show that $\gamma_{QNM} = \ln 3/ 2 \pi \sqrt{2} = 0.123637 \dots$ and $j_{min} = 1$. Although this is smaller than $\gamma_{max}$, it requires that only integer spins are counted in the sum over states (due to the value of $j_{min})$\footnote{ Interestingly, if the full calculation is done using the Meissner relation (\ref{mei-eqn}), using {\it only} integer spins, the numerical value $\gamma = 0.137727 \dots$ is obtained, again well within the bound.}. There is the advantage of linking the quantum mechanical entropy calculation to a macroscopic quantity that could be measured soon. Thus, a mechanism for ensuring that all edges of the spin network puncturing the event horizon have equal values has been suggested~\cite{dms04}.

If we take the Meissner value seriously, then one of the assumptions made to find the wave functions for the Schwarzschild interior is incorrect. Recall that we use the following ideas to obtain this bound: (1) pre-classicality to obtain sequences that are smooth far from the classical singularity; (2) the boundary condition $s_{m, n} \to 0$ as $m \to \infty$ coming from the need to match with semi-classical physics; and (3) the choice of $\delta = 2 \sqrt{3}$, based on the smallest area eigenvalue of the full theory. Since assumption (1) is necessary to make contact with the Wheeler-De Witt equation, the continuous limit of the quantum Hamiltonian, and (2) is in a similar vein, it is unlikely that they are the culprits for this inconsistency. However, it is possible the choice of $\delta$ is suspect, since it is based in part on the fiducial metric used to accomplish the quantization~\cite{ash-boj05, ash-comm}. In other words, the current method of quantization does not fix the physical value of the smallest area, since it depends on unmeasurable values - for example, the scale factor for an isotropic universe. An alternate method of quantization, where the area eigenvalues are determined {\it only} by physical quantities would result in a different value of $\delta$ (in fact, it would be a function of the triad eigenvalues), so the inconsistency may not result in that case. Work on this is ongoing at the moment~\cite{ash-comm}; it remains to be seen whether the Immirzi parameter bound $\gamma_{max}$ will change enough to include the Meissner value $\gamma_M$.

The authors appreciate the helpful comments of Abhay Ashtekar, Martin Bojowald and the other members of the Institute for Gravitational Physics and Geometry in writing this manuscript. GK is grateful for research support from the University of Massachusetts at Dartmouth, as well as the Glaser Trust.

\appendix*
\section{Exact solutions of the $\alpha_m$ recursion relation}

Since pre-classicality focuses on the behavior of the $\alpha_m$ sequences for large eigenvalues of the triad, the analysis in Section \ref{asymp-a} reflects many important features. However, the sensitivity of the solutions to the choice of initial values also needs to be studied, by examining the sequences for all values of the parameters. Generic solutions of a difference equation will feature oscillations that increase without bound; these are the sequences we label as unphysical by using the notion of pre-classicality. To weed out solutions of this type, we have to find the sequence for all values of $m$ by solving for its generating function $A(x)$; we will follow many of the steps we saw when working with the $\beta_n$ sequence. In particular, we start by defining a function $F(x)$ associated to the full generating function $A(x)$ by
\begin{equation}
\label{def-F}
F(x) = \frac{A(x) - \alpha_0 - \alpha_1 x - \alpha_2 x^2 - \alpha^3 x^3}{x^4}.
\end{equation}
Working from the recursion relation (\ref{a-true-eqn}), we arrive at the following differential equation for $F(x)$:
\begin{eqnarray}
\nonumber
& & x (x^4 - \kappa x^2 + 1) \frac{\partial F(x)}{\partial x} + F(x) (5 x^4 + \lambda x^3 - 4 \kappa x^2 - \lambda x + 3) + (\alpha_0 + \lambda \alpha_1 - 2 \kappa \alpha_2 - \lambda \alpha_3) \\
&+& (2 \alpha_1 + \lambda \alpha_2- 3 \kappa \alpha_3) x + (3 \alpha_2 + \lambda \alpha_3) x^2 + 4 \alpha_3 x^3 = 0.
\end{eqnarray}
The polynomial $(x^4 - \kappa x^2 + 1)$ factors into monomials with roots at
\begin{equation}
\label{root}
x_0 = \frac{1}{2} (\sqrt{\kappa + 2} + \sqrt{\kappa - 2}),
\end{equation}
as well as $-x_0, x_0^{-1},$ and $-x_0^{-1}$. The properties of these roots will play a role in determining whether pre-classical solutions are available for a particular $\kappa$. Already, we can see that the properties of the polynomial appearing in front of the derivative of the generating function change at the value $\kappa = 2$. We shall see that this is the point where range of pre-classical wave functions of the wave function changes as well. When we solve for the generating function $F(x)$, we find that
\begin{eqnarray}
\nonumber
F(x) &=& - x^{-3} (1 - x_0 x)^{-1/2 - \lambda/ 2 \lambda_0} (1 - x/x_0)^{-1/2 - \lambda/ 2 \lambda_0}  (1 + x_0 x)^{-1/2 + \lambda/ 2 \lambda_0}  (1 + x/x_0)^{-1/2 + \lambda/ 2 \lambda_0}  \\
\label{beast}
&\ & \times \int \biggl \lbrace x^2 \bigl[(\alpha_0 + \lambda \alpha_1 - 2 \kappa \alpha_2 - \lambda \alpha_3) + (2 \alpha_1 + \lambda \alpha_2- 3 \kappa \alpha_3) x + (3 \alpha_2 + \lambda \alpha_3) x^2 + 4 \alpha_3 x^3 \bigr] \\
\nonumber
&\ & \times (1 - x_0 z)^{-1/2 + \lambda/ 2 \lambda_0}  (1 - z/x_0)^{-1/2 + \lambda/ 2 \lambda_0}  (1 + x_0 z)^{-1/2 - \lambda/ 2 \lambda_0}  (1 + z/x_0)^{-1/2 - \lambda/ 2 \lambda_0}  \biggr \rbrace dz,
\end{eqnarray}
where the critical $\lambda$ is given by  $\lambda_0 = x_0 + x_0 ^{-1} = \sqrt{\kappa + 2}$.
Obviously, the function $F(x)$ must be the same regardless of which of the four roots we use. This can be seen by the invariance of $\lambda_0$ under the exchange $x_0 \to x_0^{-1}$, and why $\lambda_0 \to -\lambda_0$ when $x_0 \to -x_0$. The integral above is solvable in terms of Lauricella functions $F^{(4)} _D$ of four variables~\cite{ext76}, giving
\begin{eqnarray}
\nonumber
F(x) &=&  - \biggl(1 - x_0 x \biggr)^{-1/2 -\lambda/2\lambda_0} \biggl(1 - \frac{x}{x_0} \biggr)^{-1/2 - \lambda/2\lambda_0} \biggl(1 + x_0 x \biggr)^{-1/2 + \lambda/2\lambda_0} \biggl(1 + \frac{x}{x_0} \biggr)^{-1/2 + \lambda/2\lambda_0} \\
\label{full-ans}
&\ & \times \sum_{k=0} ^3 \frac{c_k x^k}{k + 3} \Laur \biggl(k+3, \frac{\lambda_0 - \lambda}{2 \lambda_0}, \frac{\lambda_0 - \lambda}{2 \lambda_0}, \frac{\lambda_0 + \lambda}{2 \lambda_0}, \frac{\lambda_0 + \lambda}{2 \lambda_0}; k+4; \frac{x}{x_0}, x_0 x, -\frac{x}{x_0}, -x_0 x\biggr).
\end{eqnarray}
Here, the coefficients $c_k$ are
\[
c_0 = - \lambda(\alpha_3 - \alpha_1) - 2 \kappa \alpha_2 + \alpha_0, \qquad
c_1 = - 3 \kappa \alpha_3 + \lambda \alpha_2  + 2 \alpha_1, \qquad
c_2 = \lambda \alpha_3 + 3 \alpha_2, \qquad
c_3 = 4 \alpha_3,
\]
and the integration constant has been chosen to be zero so that $F(0)$ gives the next value in the sequence, $\alpha_4 = c_0$, after the four initial values;  $\Laur$ is a four-variable extension of the Gaussian and Appell hypergeometric functions, defined by
\[
\Laur (a; b_1, \dots, b_4; c; z_1, \dots, z_4) = \sum_{m_1 = 0} ^\infty \dots \sum_{m_4 = 0} ^\infty \frac{(a)_{m_1 + \cdots + m_4} (b_1)_{m_1} \cdots (b_4)_{m_4} }{(c)_{m_1 + \cdots + m_4} m_1 ! \cdots m_4 !} z_1 ^{m_1} \cdots z_4 ^{m_4}.
\]
Analogous to the Appell function, the Lauricella function will converge as a series only when $|z_k| < 1$, for $k = 1, \dots, 4$. This will become important in the $\kappa > 2$ case.

First we examine the situation where $\kappa =2$. This choice of $\kappa$ simplifies the generating function greatly, since $x_0 = x_0 ^{-1} = 1$ (giving $\lambda_0 = 2$), and the Lauricella functions are reduced to Appell hypergeometric functions:
\begin{equation}
F(x) = (1 - x)^{-1 - \lambda/2} (1 + x)^{-1 + \lambda/2} \sum_{k=0} ^3 \frac{c_k x^k}{k + 3} F_1\biggl(k+3; 1 - \frac{\lambda}{2}, 1 + \frac{\lambda}{2}; k + 4; x, -x \biggr).
\end{equation}
Now we have a case similar to the $\beta_n$ sequence, and we can use the expansion of the Appell function around the point $x=1$, given by (\ref{expand}) and the exchange symmetry to see what happens at $x=-1$. Let us look first at what happens when $x = -1$. The monomial in front gives a pole of order at least one when $\lambda < 0$, and is zero when $\lambda > 2$. As we see from the first term in the Appell expansion (\ref{expand}), the hypergeometric function has singular behavior like $(1+x)^{-\lambda/2}$, giving a finite value for that term when $\lambda < 0$. Up to this point, we have the same results with the $\beta_n$ sequence -- any divergence in either the Appell function or the monomial is balanced by the reciprocal in the other. However, here we have a new wrinkle in the second term. Previously, with $\beta_n$, the Gaussian hypergeometric function in the coefficient was finite, so we did not worry about it. This is not the case here, since we have for each term in the sum $\HGeo(k+3, 1 - \lambda/2, k + 3 - \lambda/2, -x)$, which is divergent\footnote{It is important to realize that we are speaking here about divergence as a series. The hypergeometric function can be analytically continued so that a finite value is obtained at $x=-1$; this is what is done in mathematical software such as Maple and Mathematica. However, graphing the coefficients of series definition of the function will show that it has exactly the unbounded oscillatory properties that we want to eliminate.} for {\it all} values of $\lambda$~\cite{abr-ste}. This introduces the need for a relation between the $c_k$ to make the second term in the expansion around $x=-1$ finite, regardless of $\lambda$. After this, we need a second relation to insure $F(-1)$ is finite, just as in the $\beta_n$ case; it has taken two restrictions on the initial data to remove any oscillatory behavior in the sequence. When we turn to what happens at $x=1$, we find that again, we have a divergent Gaussian hypergeometric function as we approach for all $\lambda$. Thus our remaining freedom in the initial values is taken up ensuring the Appell function is not divergent. Once this is done, the first term in the expansion of the Appell function comes into play, with its singular behavior as $(1 - x)^{\lambda/2}$. Then when $\lambda < 0$, the Appell function is divergent, but together with the monomial, $F(x) \sim (1-x)^{-1}$ near $x=1$ and the sequence is bounded. When $\lambda > 0$, the monomial is singular but the hypergeometric function is non-zero (because of the second term in (\ref{expand}), now finite by selection of the $c_k$), so $F(x) \sim (1-x)^{-1- \lambda/2}$, giving an unbounded sequence.

Next we look at $\kappa \ne 2$. When $\kappa > 2$, the root $x_0$ will be a positive real number greater than one, while it is a complex number with unit modulus if $\kappa < 2$. In both cases, we can use the formula (\ref{beast}) given above, where there are potential singularities at $\pm x_0$ and $\pm x_0 ^{-1}$. Analogous to their simpler cousins the Appell functions, the parameters in the Lauricella function show there will be a pair of poles for a given value of $\lambda$, either at $x_0$ and $x_0 ^{-1}$, or else $-x_0$ and its reciprocal. For example, the hypergeometric function will have a pole at $-x_0^{-1}$ (which gives unbound oscillations) of order $1/2 - \lambda/2 \lambda_0$, i.e. when $\lambda > \lambda_0$. Similarly, there is a pole at $x = x_0^{-1}$ when $\lambda < \lambda_0$. We immediately run into a problem with the roots $\pm x_0$, however, because the Lauricella functions as a {\it series expansion} is convergent only when the variables $|z_k| < 1$. Thus when $\kappa > 2$, we have functions in the sum of the form
\[
\Laur \biggl(k+3, \frac{\lambda_0 - \lambda}{2 \lambda_0}, \frac{\lambda_0 - \lambda}{2 \lambda_0}, \frac{\lambda_0 + \lambda}{2 \lambda_0}, \frac{\lambda_0 + \lambda}{2 \lambda_0}; k+4; -1, -x_0 ^2, 1, x_0 ^2\biggr)
\]
in the sum for $F(-x_0)$. A similar result is obtained at $x = x_0$. We saw when discussing the behavior of series, in particular the series (\ref{seq-ex}) for a simple pole at $y=a$ that there may be oscillatory behavior for a finite order pole in the interval $(-\infty, -1)$, but it will eventually decay in amplitude. Here, the Lauricella functions will diverge faster than any finite power of $x$ due to the two arguments that are outside the unit box $|z_k| < 1$. We can use the expansion
\begin{eqnarray*}
\Laur (a; b_1, \dots, b_4; c; z_1, \dots, z_4) &=& \sum_{m_1=0} ^\infty \cdots \sum_{m_3 = 0} ^\infty \frac{(a)_{m_1 + m_2 + m_3} (b_1)_{m_1} (b_2)_{m_2} (b_3)_{m_3}}{(c)_{m_1 + m_2 + m_3} m_1! m_2! m_3!} \\
&\ & \times \ z_1 ^{m_1} z_2 ^{m_2} z_3 ^{m_3} \HGeo \biggl(a + \sum_{k=0} ^3 m_k; b_4; c + \sum_{k=0} ^3 m_k; z_4 \biggr)
\end{eqnarray*}
to analyze what happens at these divergent points; the choice of which of the four variables to expand around is obviously symmetric. Note that the Gaussian hypergeometric function in the sum will have the same convergence properties, regardless of the values of $m_k$, based on the real part of $(c - a - b_4)$. The result of this is that we must ensure the generating function has a finite order pole at $x = -x_0$ to avoid oscillatory sequences coming from the divergence of the Lauricella functions. This requires two relations on the initial data, since there are two of the four variables with a magnitude greater than one. The final relation comes from requiring the generating function is finite at $x= - x_0^{-1}$. Once we have gotten rid of the divergent pieces, at this value of $x$ we either have a pole in the monomial for $\lambda > - \lambda_0$, or else in the Lauricella function when $\lambda < -\lambda_0$. This is a situation similar to the $\beta_n$ sequence and the $\kappa = 2$ with a similar resolution. Specifically, $F(-x_0)$ is finite if $\lambda > -\lambda_0$, but requires a extra condition on the initial data if $\lambda < - \lambda_0$ (because the monomial is divergent but the Lauricella function is finite). It is important to note that there is not enough freedom to completely cancel out the unbounded rise of the sequence; evaluating $F(x)$ at $x = x_0$ gives the same types of divergences in the series, and one cannot impose enough conditions so their poles in the generating function are of finite order.

However, there are two particular cases for $\kappa > 2$ where it is possible to get a bounded solution -- namely, choosing $\lambda = \pm \lambda_0$ -- which are not obvious from the asymptotic analysis. If we choose $\lambda = \lambda_0$ as an example, the generating function simplifies to become
\[
F(x) = - \sum_{k=0} ^3 \frac{c_k}{k+3} \frac{F_1(k+3, 1, 1, k+4, -x/x_0, -x_0 x)}{(1 - x_0 x) (1 - x/x_0)}.
\]
The Appell function will have poles at $x = -x_0$ and $x = -x_0 ^{-1}$, as can be seen from the discussion of these functions in the last section. Since the Gaussian hypergeometric function in (\ref{expand}) is conditionally convergent at both of these values, there will only be two relations fixing the coefficients $c_k$ (this can be seen more easily when the Appell function is written in terms of logarithms and polynomials). We do not have the $\lambda$-dependent poles anymore, so this holds for any values of $\kappa > 2$. For the remaining freedom of the coefficients $c_k$, we can choose to fix it so the generating function $F(x)$ is finite at $x=x_0 ^{-1}$, to avoid unbounded solutions. This will give us a sequence that tends asymptotically to zero; the situation is similar in the case $\lambda = - \lambda_0$. These are isolated cases, however, and are not helpful in assembling a wave packet composed of a range of $\lambda$. Here we have shown it is impossible to construct generic solutions for $\kappa > 2$ if one desires the wave function to have the proper semi-classical behavior far from the classical singularity.

\begin{figure}
	\begin{minipage}[t]{0.5\linewidth}
		\centering
		\includegraphics[width=3in]{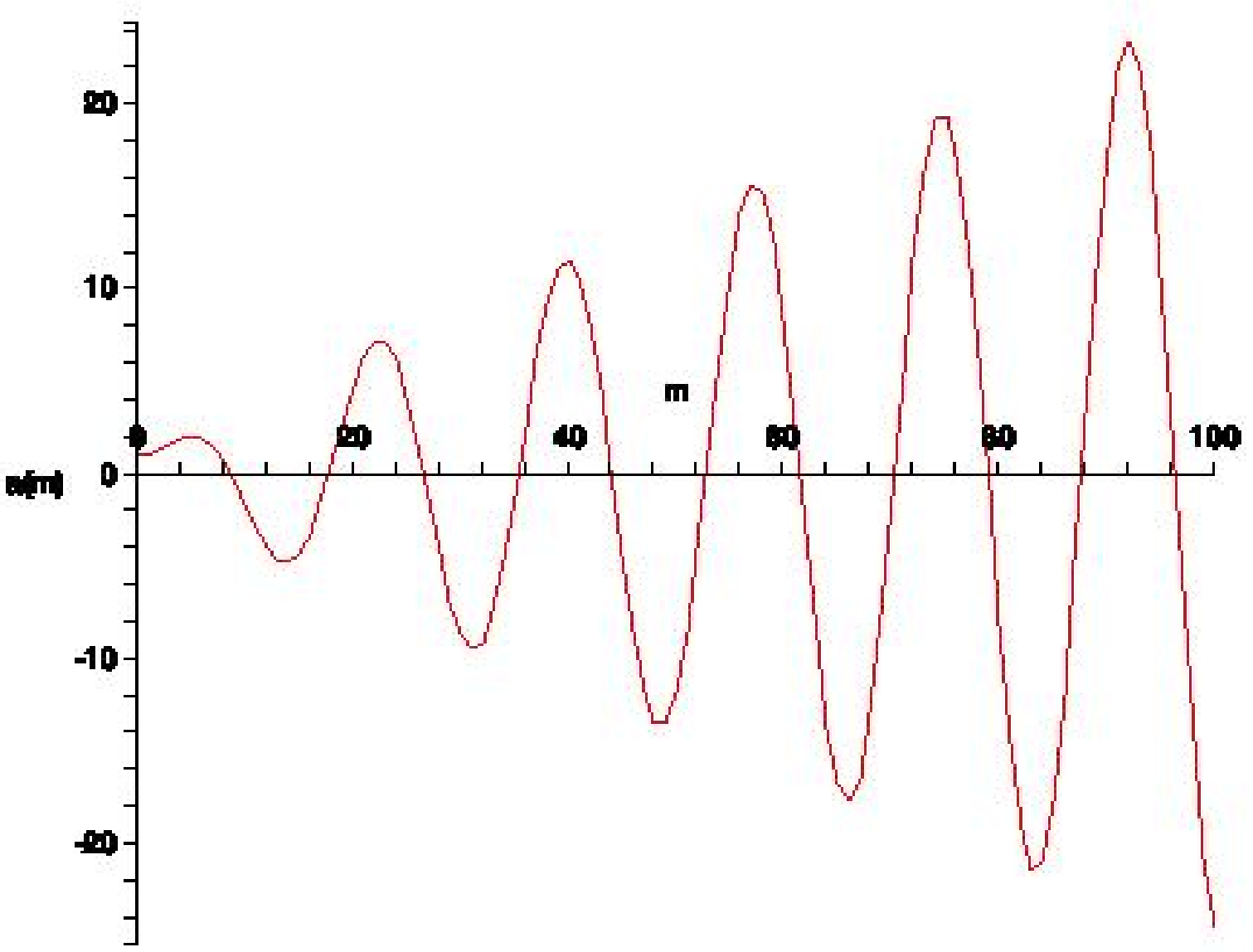}
		\caption{\label{rising}The sequence $\alpha_m$ for the case $\kappa = 3/2$ and $\lambda = 5$.}
	\end{minipage}
	\begin{minipage}[t]{0.45\linewidth}
		\centering
		\includegraphics[width=3in]{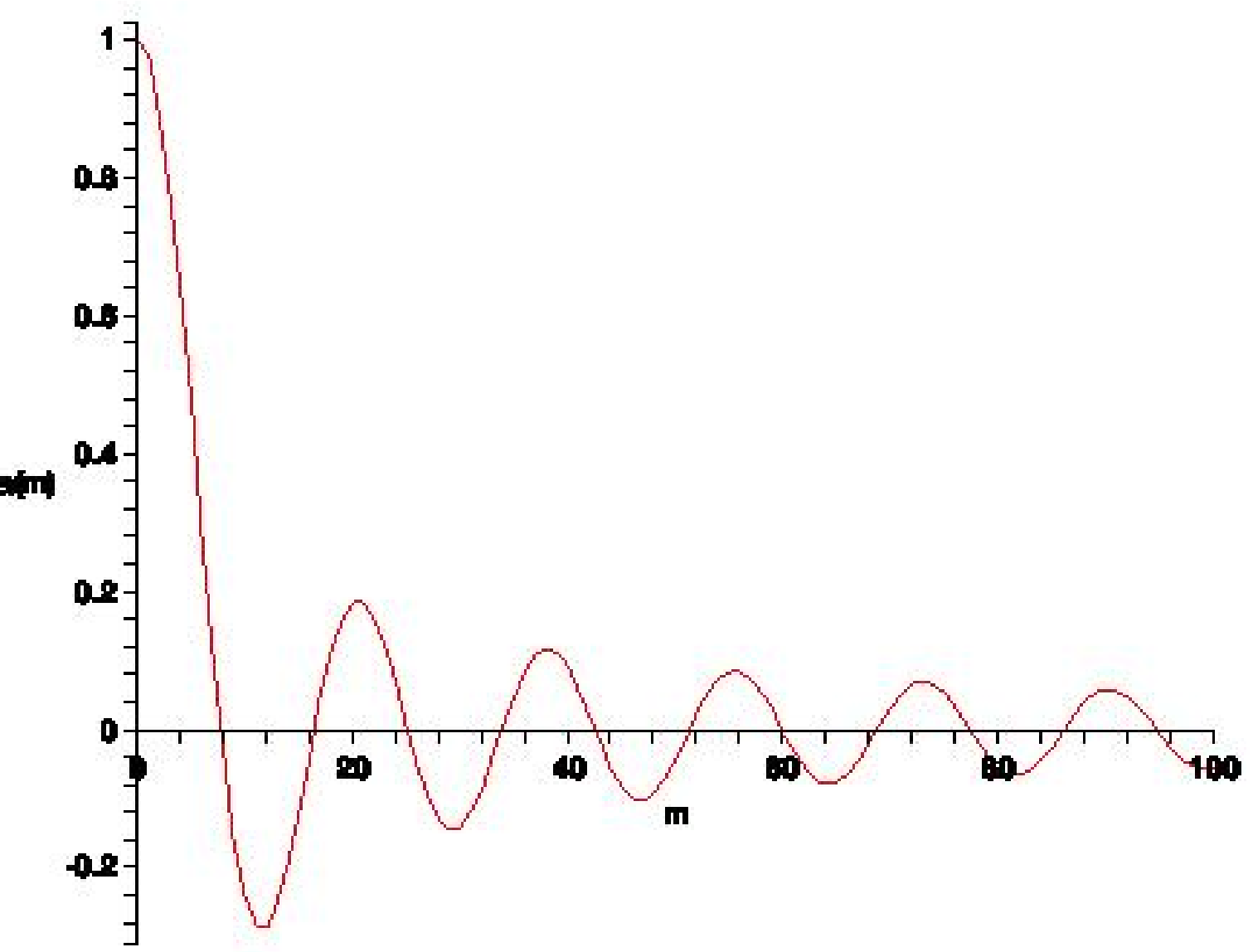}
		\caption{\label{decay} The sequence $\alpha_m$ for the case $\kappa = 3/2$ and $\lambda = -1$. Notice the period of the oscillations is the same as Figure \ref{rising}, because the value of $\kappa$ is the same.}
	\end{minipage}
\end{figure}

When $\kappa < 2$, we do not have the same difficulty, since $|x_0| = 1$. So already we have more freedom in choosing initial values for the sequence. It turns out that the initial data are completely free, because of the complex nature of the roots. Specifically, we have $x_0 = \exp (i \theta)$, where $\theta = \cos^{-1} (\sqrt{\kappa + 2}/2)$. For our range of $1 \le \kappa < 2$, $\theta$ covers the interval $0 < \theta \le \pi/6$, so none of the roots $x_0, -x_0, x_0 ^{-1}$ and $-x_0 ^{-1}$ are identical. We can see what is happening in the sequence by looking at the Taylor series of the product of two monomials, whose roots are complex conjugates. This gives
\begin{eqnarray*}
[(1 - e^{i \theta} x)(1 - e^{-i\theta} x)]^p &=& \sum_{j=0} ^\infty \sum_{k=0} ^\infty {p \choose j} {p \choose k} e^{i(j-k) \theta} (-x)^{j+k} \\
&=& \sum_{l = 0} ^\infty \bigg[ \sum_{m=0} ^l {p \choose m} {p \choose l-m} e^{i(l-2m)\theta} \biggr] x^l \\
&=& 2 \sum_{l = 0} ^\infty \bigg \lbrace \sum_{m=0} ^{[l/2]} {p \choose m} {p \choose l-m} \cos[ (l - 2m) \theta) ] \biggr \rbrace x^l \\
\end{eqnarray*}
with $[l/2]$ the greatest integer less than or equal to $l/2$. Obviously the binomials can cause the amplitude of the sequence to grow for the right range of $p$. Yet the fact we have cosine functions whose sign can change in the coefficients of $x^l$ means that we get slower oscillations in the sequence, unlike the alternating sign changes of sequences not considered physical. Since this is a generic statement, it will hold regardless of the choices of initial values, so pre-classical sequences can be found for any choice. Two particular examples are shown in Figures \ref{rising} and \ref{decay}.

\begin{thebibliography}{99}

\bibitem{lqg} A. Ashtekar and L. Lewandowski, Class. Quantum Grav. {\bf 21}, R53 (2004); C. Rovelli, {\it Quantum Gravity} (Cambridge University Press, Cambridge, 2004)

\bibitem{lqc1} A. Ashtekar, M. Bojowald and J. Lewandowski, Adv. Theor. Math. Phys. {\bf 7}, 233 (2003);

\bibitem{lqc2} M. Bojowald and H. Morales-Tecotl, Lect. Notes Phys. {\bf 646}, 412 (2004)

\bibitem{boj02} M. Bojowald, Class. Quantum Grav. {\bf 19}, 2717 (2002)

\bibitem{ash-boj05} A. Ashtekar and M. Bojowald, Class. Quantum Grav. {\bf 23}, 391 (2006)

\bibitem{mod04} L. Modesto, pre-print \url{gr-qc/0411032}

\bibitem{boj01} M. Bojowald, Phys. Rev. Lett. {\bf 87}, 121301 (2001); M. Bojowald, Gen. Rel. Grav. {\bf 35}, 1877 (2003)

\bibitem{car-kha05a} D. Cartin and G. Khanna, Phys. Rev. Lett. {\bf 94}, 111302 (2005)

\bibitem{boj-dat-hos04} M. Bojowald, G. Date and G. M. Hossain, Class. Quantum Grav. {\bf 21}, 3541 (2004)

\bibitem{car-kha-boj04} D. Cartin, G. Khanna and M. Bojowald, Class. Quantum Grav.  {\bf 21}, 4495 (2004)

\bibitem{grp-avg} D. Marolf, pre-print \url{gr-qc/0011112}

\bibitem{grp-avg-lqc} K. Noui, A. Perez and K. Vandersloot, Phys. Rev. {\bf D71}, 044025 (2005); A. Ashtekar, T. Pawlowski and P. Singh, pre-print \url{gr-qc/0602086}

\bibitem{wil93} H. Wilf, {\it Generatingfunctionology} (Academic, New York, 1993)

\bibitem{car-kha05b} D. Cartin and G. Khanna, Phys. Rev. {\bf D72}, 084008 (2005)

\bibitem{ext76} H. Exton, {\it Multiple Hypergeometric Functions and Applications} (Ellis Horwood Ltd, Chichester, 1976)

\bibitem{abr-ste} M. Abramowitz and I. Stegun, {\it Handbook of Mathematical Functions with Formulas, Graphs, and Mathematical Table} (Dover Publications, New York, 1965)

\bibitem{boj-dat-van} M. Bojowald, G. Date and K. Vandersloot, Class. Quantum Grav. {\bf 21}, 1253 (2004)

\bibitem{boj03} M. Bojowald, Class. Quantum Grav. {\bf 20}, 2595, (2003)

\bibitem{ash-bojBH} A. Ashtekar and M. Bojowald, Class. Quantum Grav. {\bf 22}, 3349 (2005)

\bibitem{mei04} K. Meissner, Class. Quantum Grav. {\bf 21}, 5245 (2004)

\bibitem{qnm} S. Hod., Phys. Rev. Lett. {\bf 81}, 4293 (1998); O. Dreyer, Phys. Rev. Lett. {\bf 90}, 081301 (2003)

\bibitem{dms04} O. Dreyer, F. Markopoulou, and L. Smolin, pre-print \url{gr-qc/0409056}

\bibitem{ash-comm} A. Ashtekar, private communication

\end{thebibliography}
\end{document}